\documentclass[prb, twocolumn]{revtex4}
\usepackage{amsmath}
\usepackage{graphicx}

\input{tcilatex}

\begin{document}

\title{Numerical Study of Spin Hall Transport in a Two Dimensional Hole Gas
System}
\author{W.Q. Chen$^{1}$, Z.Y. Weng$^{1}$, and D.N. Sheng$^{2}$}
\affiliation{$^{1}$Center for Advanced Study, Tsinghua University, Beijing 100084\\
$^{2}$Department of Physics and Astronomy, California State University,
Northridge, CA 91330}

\begin{abstract}
We present a numerical study of the spin Hall effect in a two-dimensional
hole gas (2DHG) system in the presence of disorder. We find that the spin
Hall conductance (SHC), extrapolated to the thermodynamic limit, remains
finite in a wide range of disorder strengths for a closed system on torus.
But there is no intrinsic spin Hall accumulation as induced by an external
electric field once the disorder is turned on. The latter is examined by
performing a Laughlin's Gedanken gauge experiment numerically with the
adiabatic insertion of a flux quantum in a belt-shaped sample, in which the
absence of level crossing is found under the disorder effect. Without
disorder, on the other hand, energy levels do cross each other, which
results in an oscillating spin-density-modulation at the sample boundary
after the insertion of one flux quantum in the belt-shaped system. But the
corresponding net spin transfer is only about one order of magnitude smaller
than what is expected from the bulk SHC. These apparently contradictory
results can be attributed to the violation of the spin conservation law in
such a system. We also briefly address the dissipative Fermi surface
contribution to spin polarization, which may be relevant to experimental
measurements.
\end{abstract}

\maketitle

\section{Introduction}

\label{sec:intro} Recently it has been proposed \cite{MNZ,Sinova1} that in
spin-orbit coupling (SOC)\ systems one may use an electric field to generate
transverse spin currents in the absence of external magnetic fields. It has
been argued \cite{MNZ,Sinova1} that such a spin Hall effect (SHE) is \emph{%
intrinsic}, contributed by \emph{all} the electrons below the Fermi energy,
and the corresponding spin currents are ``dissipationless''\ as in contrast
to the dissipative longitudinal charge currents which are only contributed
by the electrons close to the Fermi energy and are strongly subjected to
scattering effects.

The original proposals for the SHE are for the disorder-free cases.\cite%
{MNZ,Sinova1} In the two-dimensional (2D) electron gas described by the
Rashba model,\cite{Sinova1} it was shown based on a perturbative approach 
\cite{vertex1,vertex2} that the SHC is precisely cancelled by the vertex
correction once the disorder is turned on. On the other hand, the vertex
correction is found to vanish \cite{vertex_luttinger} for the
three-dimensional (3D) p-doped semiconductors described by the Luttinger
model \cite{MNZ} so that the SHC is still finite in the presence of weak
disorder. Numerical calculations \cite{kubo2,luttinger,kubo1} of the SHC
seem to support such perturbative results of the distinct behavior for two
models in the thermodynamic limit.

Experimentally the signatures of spin polarization have been observed
recently in 2DHG system \cite{exp1} and 3D n-doped semiconductors,\cite{exp2}
which have generated a lot of excitement concerning whether they are due to
the intrinsic SHE mentioned above or some extrinsic effect.\cite{eshe}
Bernevig and Zhang have shown that the vortex correction does vanish in the
2DHG \cite{2dhg} and 3D n-doped semiconductor \cite{ngaas} systems.
Furthermore, the mesoscopic SHE in the 2DHG is also found to be present
based on the nonequilibrium Green function method \cite{ne2dhg} and
Landauer-B\"{u}ttiker formula with attached leads,\cite{lb2dhg} similar to
(with larger magnitude than) the mesoscopic SHE found in the 2D Rashba model.%
\cite{LB1, LB3} But it is still unclear whether the bulk SHE in the 2DHG can
survive in the thermodynamic limit beyond the perturbative approach.

In this paper, we perform numerical calculations for the 2DHG in the
presence of disorder. We first show that the bulk SHC calculated from the
Kubo formula is indeed robust against the disorder in extrapolation to the
thermodynamic limit, which is consistent with the vertex correction
calculation.\cite{2dhg} It is also similar to the behavior for the Luttinger
model,\cite{luttinger} but is in opposite to that\ of the 2D Rashba model.%
\cite{kubo2} But when we perform a Laughlin's Gedanken ``gauge experiment''\
on a belt-shaped sample to probe the spin transfer/accumulation due to the
SHE, we obtain a null result due to the anticrossing between energy levels,
which is quite similar to what has been previously seen in the 2D Rashba
model.\cite{kubo2} We point out that the absence of edge states in the 2DHG
system causes the general level repulsion with the turn on of disorder,
which leads to the disappearance of the net spin Hall accumulation in an
open system. Furthermore, at zero disorder case, energy levels do cross each
other and we find an oscillating spin-density-modulation at the sample
boundary after an adiabatic insertion of a flux quantum in the
above-mentioned Laughlin's gauge experiment. However, the corresponding net
spin transfer is only about one order of magnitude smaller than what is
expected from the calculated bulk SHC. We discuss the ``conflicting''\
results of a finite SHC but without an intrinsic spin accumulation in the
2DHG, and point out that the underlying reason can be attributed to the
violation of the spin conservation law in such a SOC system, where the SHC
is no longer an unambiguous quantity for describing the spin transport. We
also address the addition dissipative Fermi surface contribution to spin
polarization, which may be relevant to experimental measurement.\cite%
{exp1,exp2}

The remainder of the paper is organized as follows. In Sec. \ref{sec:shc},
we numerically compute the SHC through the Kubo formula in a tight-binding
model of the 2DHG system at different sample sizes and disorder strengths,
and perform finite-size scaling analysis. In Secs. \ref{sec:gelc}, \ref%
{sec:st} and \ref{sec:nr}, we perform a Laughlin's gauge experiment
numerically to determine the intrinsic spin transfer/accumulation due to the
adiabatic insertion of a magnetic flux quantum. And in Sec. \ref{sec:dis},
we compare the results with a system in a perpendicular external magnetic
field and demonstrate the relationship between the intrinsic spin transfer
(accumulation) and the existence of edge states. Finally, a summary is given
in Sec. V.

\section{Spin Hall Conductance}

\label{sec:shc}

\subsection{Hamiltonian}

\begin{figure}[t]
\resizebox{80mm}{!}{    \includegraphics{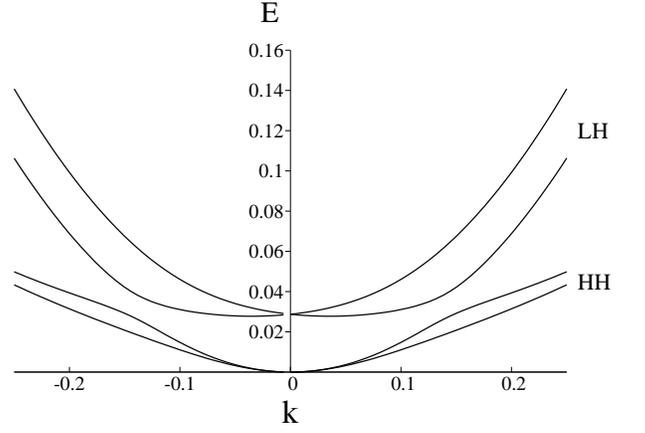} } \centering
\caption{Band structure for a clean 2DHG system determined by Eq.(  \protect
\ref{H}) (with $t=1)$. The HH and LH denote the heavy hole band and light
hole band, respectively.}
\label{fig:band}
\end{figure}

We start with the 2DHG Hamiltonian proposed by Bernevig and Zhang\cite{2dhg} 
\begin{equation}
H=(\gamma _{1}+\frac{5}{2}\gamma _{2})\frac{k^{2}}{2m}-\frac{\gamma _{2}}{m}(%
\vec{k}\cdot \vec{S})^{2}+\alpha (\vec{S}\times \vec{k})\cdot \hat{z}
\label{eq:con_ham}
\end{equation}%
For the convenience of the following numerical study, we further convert
this continuum Hamiltonian into a tight-binding version on the square
lattice. This may be realized by making the replacement $k_{\nu }\rightarrow
\sin k_{\nu }$ and $k_{\nu }^{2}\rightarrow 2(1-\cos k_{\nu })$. Two models
are apparently equivalent near the band bottom where $k_{\nu }\rightarrow 0$%
. The resulting Hamiltonian reads 
\begin{eqnarray}
H &=&-t\sum_{\langle ij\rangle }(c_{i}^{\dagger }c_{j}+c_{j}^{\dagger
}c_{i})+V_{L}\sum_{i\nu }(c_{i}^{\dagger }S_{\nu }^{2}c_{i+\hat{\nu}}+H.c.) 
\notag \\
&+&\frac{V_{L}}{4}\left( \sum_{i}c_{i}^{\dagger }\left\{ S_{x},S_{y}\right\}
(c_{i+\hat{x}+\hat{y}}-c_{i+\hat{x}-\hat{y}})+H.c.\right)  \notag \\
&+&V_{R}\left( i\sum_{i}c_{i}^{\dagger }S_{y}c_{i+\hat{x}}-i\sum_{i}c_{i}^{%
\dagger }S_{x}c_{i+\hat{y}}+H.c.\right)  \notag \\
&-&2V_{L}\sum_{i}c_{i}^{\dagger }\left[ (S_{x}^{2}+S_{y}^{2})-(\frac{9}{4}%
-S_{z}^{2})\langle k_{z}^{2}\rangle \right] c_{i}  \notag \\
&&+4t\sum_{i}c_{i}^{\dagger }c_{i}\text{ }+\epsilon _{i}
\sum_{i}c_{i}^{\dagger }c_{i}\text{ }  \label{H}
\end{eqnarray}%
where the electron annihilation operator $c_{i}$ has four components
characterized by the ``spin'' index $S_{z}=\frac{3}{2},\frac{1}{2},-\frac{1}{%
2},-\frac{3}{2},$ respectively, and $i+\hat{\nu}$ ($\hat{\nu}=\hat{x},\hat{y}
$) denote the nearest-neighbors of site $i$. Here $t$ is the
nearest-neighbor hopping integral defined as 
\begin{equation}
t=\frac{\hbar ^{2}}{2ma_{0}^{2}}(\gamma _{1}+\frac{5}{2}\gamma _{2}),
\end{equation}%
where $m$ is the mass of the electron and $a_{0}$ is the lattice constant. $%
V_{L}$ and $V_{R}$ represents the strengths of the Luttinger-type and
Rashba-type spin-orbital couplings, respectively: 
\begin{eqnarray}
V_{L} &\equiv &\frac{2\gamma _{2}}{\gamma _{1}+\frac{5}{2}\gamma _{2}}t 
\notag \\
V_{R} &\equiv &\frac{\alpha }{\gamma _{1}+\frac{5}{2}\gamma _{2}}t.  \notag
\end{eqnarray}%
The parameters $\gamma _{1}$ and $\gamma _{2}$ in GaAs are given by $\gamma
_{1}=6.92$ and $\gamma _{2}=2.1$,\cite{haug} so we have $t\sim 1.45eV$ and $%
V_{L}\sim 0.345t$. We shall choose $V_{R}=0.02t$ for the Rashba-type
coupling. By assuming that the 2DHG is confined in a well of the thickness $%
a\sim 8.7$ nm, one approximately has $\langle k_{z}\rangle =0$ and $\langle
k_{z}^{2}\rangle \sim \left( \frac{\pi }{a/a_{0}}\right) ^{2}$, where $%
a_{0}\sim $ $0.565$ nm for GaAs. The corresponding energy spectrum for a
clean sample ($\epsilon_{i}=0$) is shown in Fig. \ref{fig:band}. Due to the
confinement in the $\hat{z}$ direction, a gap with $0.029t\sim 0.042eV$
opens up between the heavy hole (HH) band and the light hole (LH) band at $%
\Gamma $ point (see Fig. \ref{fig:band}). For simplicity, we shall set $t=1$
in the rest of paper. Finally, $\epsilon _{i}$ in Eq.(\ref{H}) accounts for
the on-site and spin-independent disorder strength, which is randomly
distributed within $[-W/2,W/2]$.

\subsection{\protect\bigskip Spin Hall conductance}

The SHC is defined by the Kubo formula\cite{kubo0} 
\begin{equation}
\sigma _{SH}=\frac{2}{N}\left\langle \mathrm{Im}\sum_{E_{n}<E_{f}<E_{m}}%
\frac{\langle n|j_{x}^{z\mathrm{spin}}|m\rangle \langle m|j_{y}|n\rangle }{%
(E_{m}-E_{n})^{2}}\right\rangle ,  \label{eq:kubo}
\end{equation}%
where $N$ is the total number of lattice sites, $E_{f}$ denotes the Fermi
energy, $E_{m}$ ($E_{n}$) is the eigen-energy, and $\langle ...\rangle $ is
averaged over all disorder configurations. The charge current operator $%
j_{\mu }$ and spin current operator $j_{\mu }^{\nu \mathrm{spin}}$ are
defined as 
\begin{eqnarray}
j_{\mu } &\equiv &ev_{\mu },  \notag \\
j_{\mu }^{\nu \mathrm{spin}} &\equiv &\frac{1}{2}\{v_{\mu },S_{\nu }\}, 
\notag
\end{eqnarray}%
where the velocity operator $\mathbf{v}$ is the conjugate operator of the
position operator $\mathbf{R}\equiv \sum_{i\sigma }\mathbf{r}_{i}n_{i\sigma
} $ ($n_{i\sigma }$ is the electron number operator at site $i$ with spin
index $\sigma $), is defined by the standard relation 
\begin{equation}
v_{\mu }=\frac{i}{\hbar }[H,R_{\mu }]\text{ \ }.
\end{equation}

\begin{figure}[t]
\resizebox{90mm}{!}{    \includegraphics{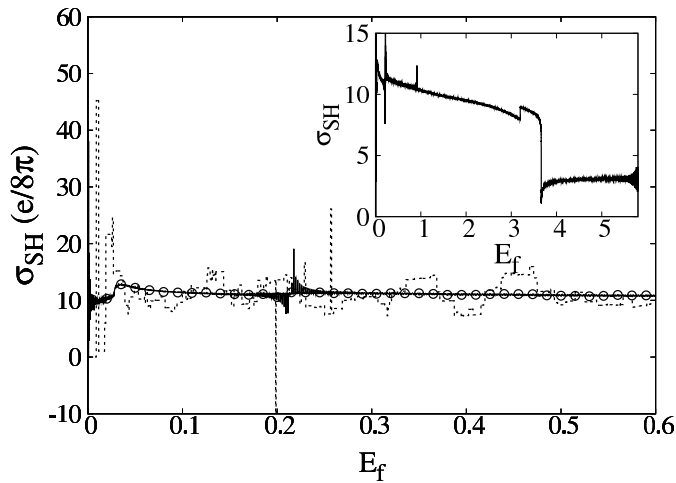} } \centering
\caption{$\protect\sigma_{SH}$ vs $E_f$ at $W = 0$. The inset shows the
whole Fermi energy region and the main panel shows the details around the
band bottom. The solid line is at $1000 \times 1000$ lattice, the dashed
line is at $24 \times 24$ lattice, and the open circles are at $24 \times 24$
lattice with averaged over 1000 BCs.}
\label{fig:pure}
\end{figure}

For a pure sample ($\epsilon _{i}=0$), the Hamiltonian can be easily
diagonalized in momentum space, and we can determine\ $\sigma _{SH}$ vs $%
E_{f}$ for a very large lattice ($N=L^{2}$ with $L=1000$) as shown in the
inset of Fig. \ref{fig:pure}. Since the tight-binding model is a good
approximation of the original model only when $E_{f}$ is near the band
bottom, we shall focus on the regime of $0<E_{f}<0.6$ as represented in the
main panel of Fig. \ref{fig:pure} by the solid curve. One can see that $%
\sigma _{SH}$ in this region is a little bit larger than $10$ $\mathrm{%
e/8\pi }$ and very flat as a function of $E_{f}$ except for that within the
gap ($\sim 0.029)$ between the HH and LH bands.

Because the momentum is no longer a good quantum number in the presence of
disorder, Eq.(\ref{H}) can not be diagonalized in the momentum space when $%
\epsilon _{i}\neq 0$. In the latter case, the maximal lattice size reachable
by the numerical calculation is usually much smaller than $L=1000$ (around $%
L=20-30$). To check what happens at smaller lattices, we compute $\sigma
_{SH}$ on an $L=24$ lattice in the pure system with a periodic boundary
condition (BC) and the result is illustrated by the dotted curve in Fig. \ref%
{fig:pure}. The corresponding $\sigma _{SH}$ exhibits very large
fluctuations whose magnitude can be several times larger than the converged
value at $L=1000$. To remove such finite-size fluctuations, we introduce an
average over different BCs. A general (twisted) BC is given by 
\begin{eqnarray}
\psi (x+L,y) &=&e^{i\theta _{x}}\psi (x,y)  \notag  \label{eq:tbc} \\
\psi (x,y+L) &=&e^{i\theta _{y}}\psi (x,y)
\end{eqnarray}%
where $\theta _{\mu }$'s are defined within $[0,2\pi ]$ with $\theta _{\mu
}=0$ or $2\pi $ ($\mu =x,y)$ corresponding to the periodic BC. Since the SHC
should not be sensitive to the BCs in the thermodynamic limit, in principle, 
$\sigma _{SH}$ in Eq. \eqref{eq:kubo} can be always defined as averaged over
both disorder and twisted BCs in such a limit. For a clean system, the
averaged $\sigma _{SH}$ at $L=24$ over $1000$ randomly generated BCs is
shown by the open circles in Fig. \ref{fig:pure}. One can clearly see that
the large fluctuations in $\sigma _{SH}$ at $L=24$ (dotted curve) have been
significantly reduced after the BC averaging, and the result (open circles)
coincides with $\sigma _{SH}$ obtained at $L=1000$ (solid line) very well.
It indicates that the finite-size effect at smaller size ($L=24$) can be
properly removed by the BC average. As a consequence, the BC averaged SHC
behaves much smoother than at a fixed BC, and is thus more suitable for a
finite-size scaling analysis. We note that the BC average is also required
by the gauge-invariant condition for SHC in a finite size system. When we
apply an electric field through a time-dependent vector potential, it acts
as a generalized boundary phase evolving with time (see below). Thus the SHC
averaged over time is gauge-invariant and equivalent to the BC averaged SHC.%
\cite{kubo2} %In the large system size limit, SHC becomes independed of the
%BC and thus the BC average is no longer needed.
Such a numerical method has been previously used in the study of the 2D
Rashba model and 3D Luttinger model.\cite{kubo2,luttinger} Similar boundary
phase averaged charge Hall conductance can be also related to a topological
invariant Chern number.\cite{kubo0}

\begin{figure}[t]
\resizebox{60mm}{!}{    \includegraphics{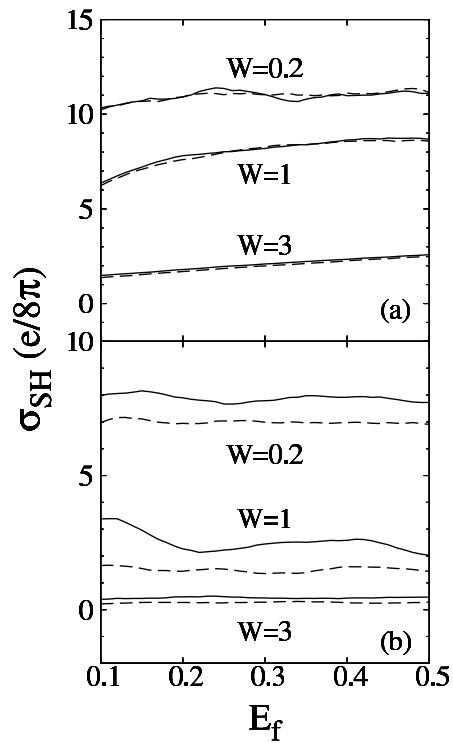} } \centering
\caption{$\protect\sigma _{SH}$ vs. $E_{f}$ at $W=0.2,$ $1,$ and $3$. The
solid curves are for $L=16$, and the dashed ones for $L=24$. (a) is for the
2DHG system and (b) is for the spin-3/2 Rashba model by setting $V_L=0$.}
\label{fig:impurity}
\end{figure}

Now we use the above method to study the sample-size dependence of $\sigma
_{SH}$ at $W=0.2$ for weak disorder strength; $W=1$ for intermediate
disorder; and $W=3$ for strong disorder. The calculated $\sigma _{SH}$ is
presented in Fig. \ref{fig:impurity} (a) for two sample sizes: $L=16$
[averaged over 20000 configurations (solid line)] and $L=24$ [averaged over
5000 configurations (dashed line)]. The size-dependence of $\sigma _{SH}$ is
very weak for all three cases at different $E_{f}$'s. For comparison, we
have also done a similar calculation for a ``spin-3/2 (linear) Rashba
model'', which corresponds to Eq.~(\ref{H}) at $V_{L}=0$ and $V_{R}=0.5$.
SHC at $W=0.2,$ $1,$ and $3,$ with $L=16$ and $24,$ respectively, are shown
in Fig. \ref{fig:impurity}(b), which clearly decrease with the increase of
the sample length, similar to the result in the spin-1/2 Rashba model.\cite%
{kubo2} 
%%where a scaling analysis shows $\sigma _{SH}\rightarrow 0$ when the sample
%%size goes to infinity.

\begin{figure}[t]
\resizebox{80mm}{!}{    \includegraphics{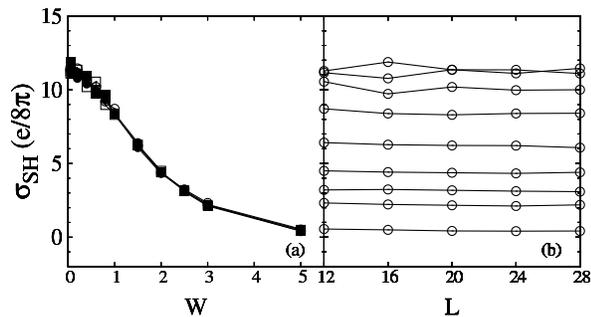} } \centering
\caption{$W$ and sample-size dependence of $\protect\sigma _{SH}$ with $%
E_{f} $ averaged within the interval $[0.3,0.4]$. (a) $\protect\sigma _{SH}$
vs. $W $. The open squares are for $L=12,$ close squares for $L=16$, open
circles for $L=20$, and close circles for $L=24$. (b) $\protect\sigma _{SH}$
vs $L$. The lines from top to bottom are for $W=0.05,$ $0.2,$ $0.6,$ $1,$ $%
1.5,$ $2,$ $2.5,$ $3,$ and $5$, respectively.}
\label{fig:dep}
\end{figure}

To further investigate the effects of the lattice size and disorder, we
calculate the averaged $\sigma _{SH}$ within a chosen energy interval $%
[0.3,0.4]$ at various disorder strengths and sample sizes as shown in Fig. %
\ref{fig:dep}. $\sigma _{SH}$ is almost independent of sample sizes and
reduces monotonically with increasing $W$, which is different from the
spin-1/2 Rashba model but similar to the behavior in the 3D Luttinger model.%
\cite{luttinger} In the weak disorder limit, $\sigma _{SH}$ approaches to
the value slightly larger than $10$ $\mathrm{e/8\pi }$, which is quite close
to the value ($\sim 9e/8\pi $) obtained \cite{loss} in an analytic
calculation in a different model for 2DHG system. Thus we establish that SHC
is finite extrapolating to thermodynamic limit until the disorder strength
reaches a critical value $W_{c}=5.0$.

\section{Intrinsic Spin Transfer and Accumulation}

\label{sec:sa}

In the last section, we have shown the numerical evidence that a finite $%
\sigma _{SH}$ can survive in the presence of disorder with the sample size
being extrapolated to the thermodynamic limit. This is in sharp contrast to
the 2D Rashba model in which $\sigma _{SH}$ vanishes in the thermodynamic
limit with the turn on of very weak disorder strength.\cite{kubo2} Such a
distinct behavior in $\sigma _{SH}$ is consistent with the perturbation
theories in which the vertex correction also results in opposite conclusions
on the fate of $\sigma _{SH}$ in two models, as mentioned in the
Introduction.

In general, a finite conductance should lead to an accumulation of current
carriers at the edges (open boundaries) of the sample according to the
current conservation law. However, in the SOC system the spin current
conservation law does not exist as the spin is not conserved. Thus the
relation between the SHC and the spin accumulation is not straightforwardly
present.

In this section, we will try to probe the spin transfer or accumulation
effect directly by performing numerically a Laughlin's gauge experiment,
which was first used in the IQHE system \cite{laughlin, halperin} to explain
why an integer number of charges can be transported across the sample in the
transverse direction when a longitudinal electric field is applied via an
adiabatic flux insertion process. This method has been recently generalized
to the SOC system by Sheng \textit{et al.}\cite{kubo2} in the study of the
SHE in the 2D Rashba model.

\subsection{Gauge experiment and level crossing}

\label{sec:gelc}

\begin{figure}[t]
\resizebox{90mm}{!}{    \includegraphics{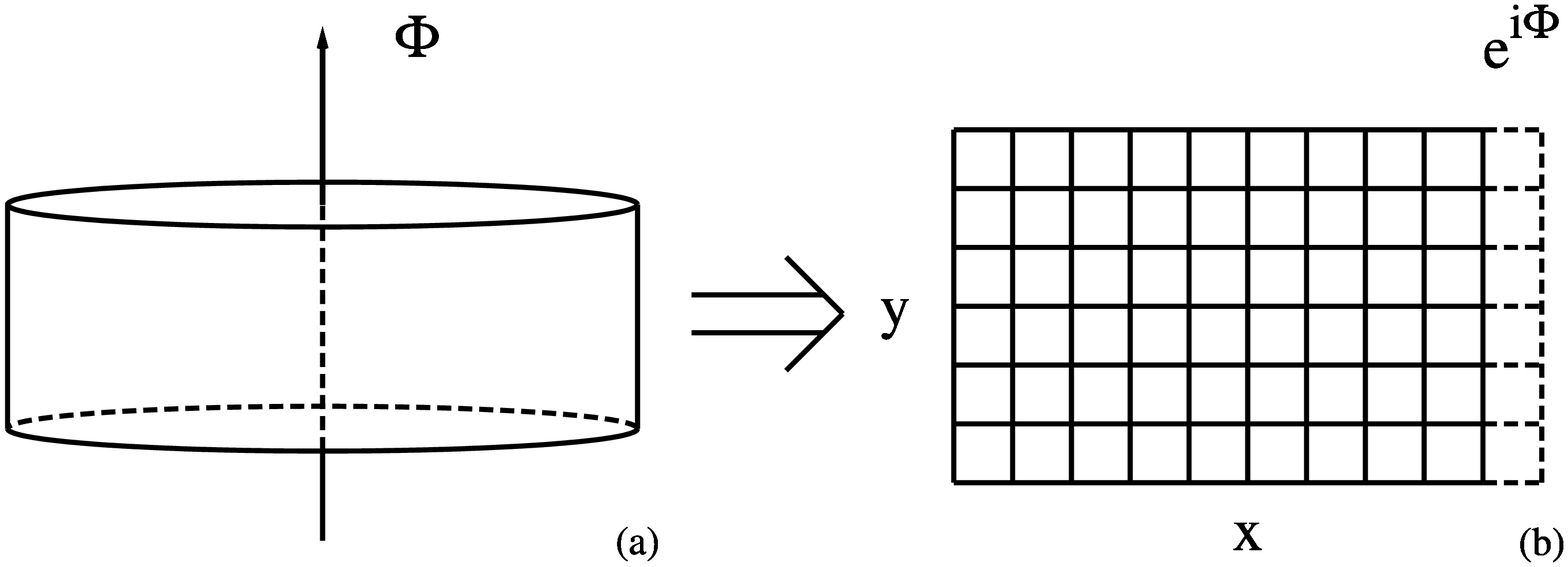} } \centering
\caption{(a) The geometry of the sample in the gauge experiment. It is like
a belt enclosing a flux in the center. (b) The implementation of (a) in the
numerical calculation. The dashed line with a phase factor $e^{i\Phi }$
along the $\hat{x}$ axis means a twist BC, which represents the flux in (a).}
\label{fig:geometry}
\end{figure}

The numerical gauge experiment will be performed on a belt-shaped sample
with a magnetic flux $\Phi $ adiabatically inserted at the center of the
ribbon [see Fig. \ref{fig:geometry}(a)]. Such a belt-shaped geometry with
the magnetic flux can be implemented in a square lattice sample shown in
Fig. \ref{fig:geometry}(b) by imposing a twisted BC along the $\hat{x}$
direction and an open BC along the $\hat{y}$ direction. Namely, with a gauge
transformation, one may impose the flux $\Phi $ by changing the BC along $%
\hat{x}$ direction from the periodic BC to a twist BC with the phase twist $%
\theta _{x}=\Phi $ indicated by the dashed line and the phase factor $%
e^{i\Phi }$ in Fig. \ref{fig:geometry}(b).

\begin{figure}[t]
\resizebox{80mm}{!}{    \includegraphics{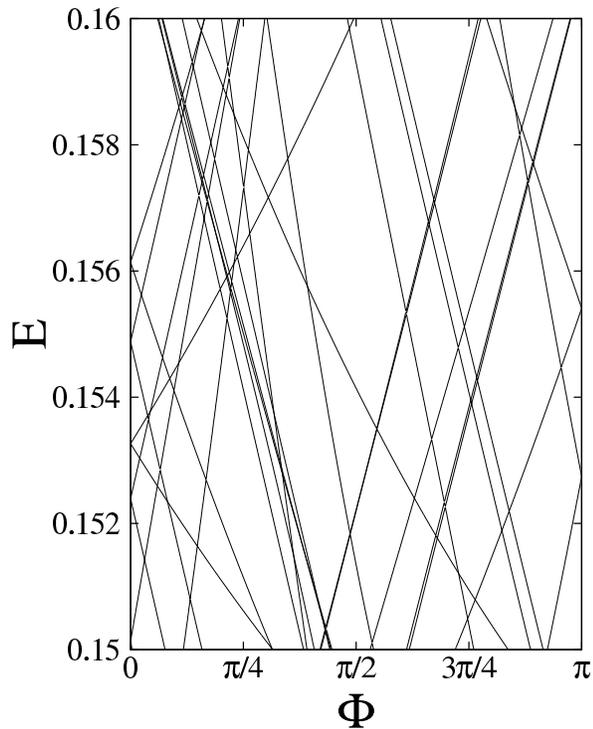} } \centering
\caption{The evolution of the single electron state energy with$\ \Phi $ for
the 2DHG on a $32\times 32$ clean sample ($W=0$).}
\label{fig:level_pure}
\end{figure}

Numerically the energy spectrum of single electron states at a fix $\Phi $
can be calculated in the geometry of Fig. \ref{fig:geometry}(b). Fig. \ref%
{fig:level_pure} shows such calculated energy spectra as a function of $\Phi 
$ in the pure system. Note that the system is equivalent at $\Phi =0$ and $%
\Phi =2\pi $ as the Hamiltonian is periodic $H(\Phi =0)=H(\Phi =2\pi ),$ and
the energy spectrum is symmetric between $0\leq \Phi \leq \pi $ and $\pi
\leq \Phi \leq 2\pi $ due to a time-reversal symmetry, such that only the
first half is plotted in Fig. \ref{fig:level_pure}.

One may imagine changing the flux $\Phi $ adiabatically from $0$ to $2\pi $
to follow the corresponding change of the system. Generally there are two
possibilities for the evolution of the ground state with the electrons
filled below the Fermi energy $E_{f}$. If each single electron state always
anti-crosses with other states (level repulsion), it will eventually return
to the original state after the insertion of one flux quantum of $\Phi =2\pi 
$ sufficiently slow (i.e., adiabatically such that there is no quantum
transition to other levels during this operation); On the other hand, if a
level crossing occurs for single electron states during the adiabatical
increase of $\Phi ,$ then the original ground state generally will evolve
into a different state through the change of the single state occupation
number, even though the single electron energy spectrum must remain the same
after the insertion of one flux quantum.

A slow increase of the flux $\Phi $ is equivalent to applying a weak
electric field along the $\hat{x}$ direction:%
\begin{equation}
\epsilon _{x}=-\frac{1}{L_{x}}\frac{\partial \Phi }{\partial t}.
\label{electric}
\end{equation}%
Then the \emph{intrinsic} charge/spin Hall effect due to an electric field
in belt-shaped system will lead to the transfer of charge/spin along the $%
\hat{y}$-axis during the flux insertion. If there is to be charge/spin
accumulation at the open edges [Fig. \ref{fig:geometry}(a)], it means that
the ground state should not return to the original one after the adiabatic
flux insertion, which is the essence of Laughlin's gauge experiment for the
IQHE system. Therefore, the level crossing of single electron states serves
as a necessary condition for the existence of \emph{intrinsic} spin
transfer/accumulation for a SHE system. As clearly shown in Fig. \ref%
{fig:level_pure}, which is determined on an $L=32$ lattice, the single
electron energy levels do simply cross in the pure 2DHG system. The
resulting net spin transfer will be analyzed in the following section. 
%%question if thus This property is thus consistent with the intrinsic
%%SHE in the pure sample discussed above.

\subsection{Spin Transfer at $W = 0$}

\label{sec:st}

\begin{figure}[t]
\resizebox{80mm}{!}{    \includegraphics{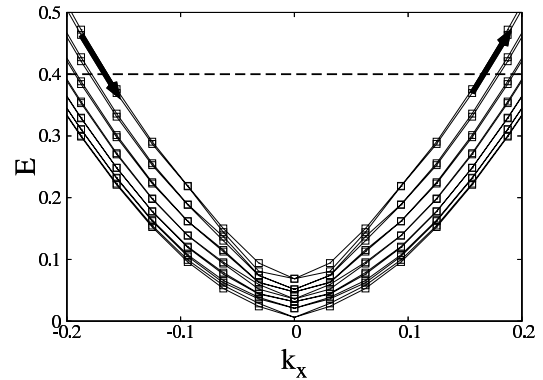} } \centering
\caption{The miniband structure of a $32 \times 32$ clean sample. Two arrows
represent the evolution of two states by tuning the parameter $\Phi $, which
start from the beginning at $\Phi =0$ and arrive at the end of the arrows at 
$\Phi =2\protect\pi $ (see the text for details). The dashed line indicates
the Fermi energy.}
\label{fig:band_o}
\end{figure}

%As a necessary condition of the spin transport, we have shown the level
%crossing in the pure sample in last section.
In this section, we calculate the spin transfer induced by tuning the flux $%
\Phi $ from $0$ to $2\pi $ in the pure case ($\epsilon_{i}=0$). Starting
from the periodic BC ($\Phi =0$), the momentum $k_x$ along $\hat{x}$
direction is a good quantum number. And due to the open boundary along the $%
\hat{y}$ direction, the electron states form a mini-band structure which is
shown in Fig. \ref{fig:band_o} and can be denoted as $|k_{x},n\rangle $,
where $n$ is the index of the mini-bands. If we choose an arbitrary Fermi
energy $E_{f}$, the ground state corresponds to that all the single particle
states below $E_{f}$ are occupied, whereas unoccupied above $E_{f}$. So the
spin density along $\hat{y}$ direction (labelled by $y\equiv j_{y}$) at the
initial state can be defined by 
\begin{equation}
S_{z}^{i}(y)\equiv \sum_{E_{k_{x},n}<E_{f}}\langle
k_{x},n|\sum_{j_{x}}c_{j}^{+}S_{z}c_{j}|k_{x},n\rangle .
\end{equation}%
When one inserts the flux adiabatically from zero to $\Phi $, the state $%
|k_{x},n\rangle $ will evolve to $|P_{x}=k_{x}+\Phi /L_{x},n\rangle $, For $%
k_{x}>0$ and $<0$, the corresponding energies of the states will increase
and decrease, respectively, which is illustrated by arrows in Fig. \ref%
{fig:band_o}. Thus one finds a simple level crossing for two levels close to
each other with opposite signs of $k_{x}$, which represents the level
crossing shown in Fig. \ref{fig:level_pure} for pure system due to no mixing
term between different states in the Hamiltonian. After the insertion of one
flux quantum, the state $|k_{x},n\rangle $ evolves into a new state $%
|k_{x}+2\pi /L_{x},n\rangle $ at $\Phi =2\pi $. Then the spin density at the
final state is given by 
\begin{equation}
S_{z}^{f}(y)=\sum_{E_{k_{x},n}<E_{f}}\langle k_{x}+\frac{2\pi }{L_{x}}%
,n|\sum_{j_{x}}c_{j}^{+}S_{z}^{j}c_{j}|k_{x}+\frac{2\pi }{L_{x}},n\rangle .
\end{equation}%
The resulting change in the spin-density due to threading a flux quantum is $%
\Delta S_{z}(y)=S_{z}^{f}(y)-S_{z}^{i}(y)$.

\begin{figure}[t]
\resizebox{90mm}{!}{    \includegraphics{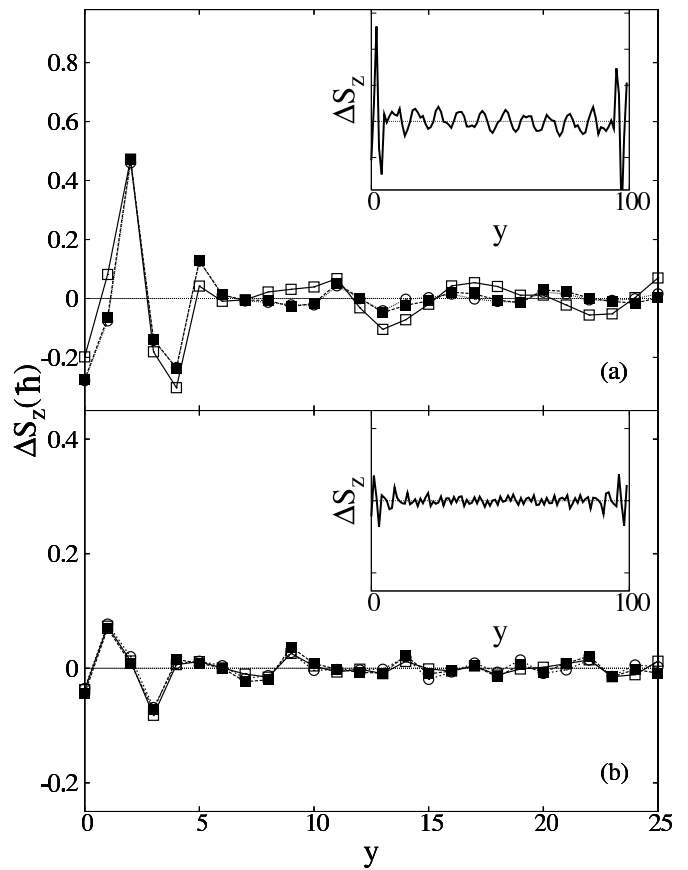} } \centering
\caption{The spin density changes between the state at the beginning and the
end of the procedure of increasing the flux from $0$ to $2 \protect\pi$ in
the pure samples. The main panel shows the data around one edge for various
lattice sizes, while the inset shows the whole result on a $500 \times 100$
lattice. The Fermi energy is fixed at $E_f = 0.4$. (a) is for the 2DHG; The
solid lines with open squares are for $1000 \times 100$ lattice, the dashed
lines with close squares are for $500 \times 500$ lattice, and the dotted
lines with open circles are for $1000 \times 1000$ lattice. (b) is for the
2D Rashba model; The solid lines with open squares are for $100 \times 100$
lattice, the dashed lines with close squares are for $500 \times 100$
lattice, and the dotted lines with open circles are for $1000 \times 1000$
lattice.}
\label{fig:spin_dis}
\end{figure}

$\Delta S_{z}(y)$ in a pure 2DHG with $500\times 100$ lattice is shown in
the inset of Fig. \ref{fig:spin_dis}(a), where the Fermi energy is $%
E_{f}=0.4 $. We recall that due to the time-reversal invariance, $%
S_{z}^{i}(y)=0$ everywhere. After adding the $2\pi $ flux quantum, the spin
density shows some strong peaks at the boundaries whereas with a much weaker
magnitude in the bulk. 
%To see more clearly and consider the lattice size effect,
We show the spin-wave like modulation in $\Delta S_{z}(y)$ near one edge of
the belt-shaped sample at various lattice sizes in the main panel of Fig. %
\ref{fig:spin_dis}(a). 
%%It is clearly that the size dependence of $\Delta S_z (y)$  is different
%%at the edge and at the edge and in the bulk is different.
Here $\Delta S_{z}(y)$ is almost sample-size independent at the edge,
whereas it reduces monotonically with the sample size in the bulk. So in the
thermodynamic limit, there should be only the spin density modulation
present near the edge. Similar results with smaller magnitude are also
obtained for an electron spin-$1/2$ Rashba model as shown in Fig. \ref%
{fig:spin_dis}(b) for comparison. Such a spin density modulation does
support a net spin transfer along the $\hat{y}$ direction, accompanying an
electric field generated along the $\hat{x}$ direction due to the inserting
flux according to Eq. (\ref{electric}), but the magnitude of the total
(integrated) $\Delta S_{z}\sim 0.2\hbar$ at one edge is much less than one
spin quantum, which is in sharp contrast to the large bulk SHC (the
calculated $\sigma _{SH}\sim 11 \mathrm{e/8\pi }$, that suggests a total
spin transfer of $5.5(\hbar/2)$ spin quanta). This large discrepancy between
two approaches indicates that the SHC is no longer an unambiguous quantity
for measuring the spin transport even in the pure case. In the following we
further explore the disorder case.

\subsection{Disorder effect}

\label{sec:nr}

\begin{figure}[t]
\resizebox{80mm}{!}{    \includegraphics{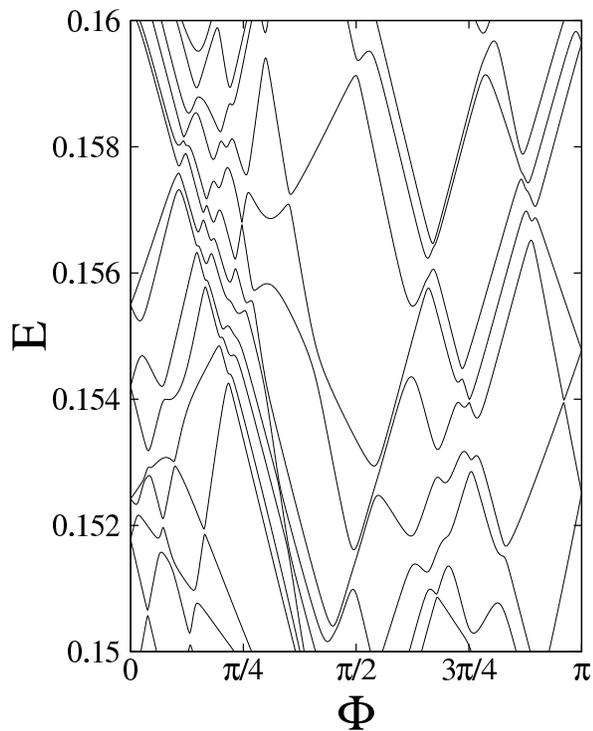} } \centering
\caption{The evolution of the single electron states with$\ \Phi $ for the
2DHG on $32\times 32$ lattice for a disordered sample with $W=0.05$. }
\label{fig:level}
\end{figure}

Now we consider the disorders effect. 
%%The on-site disorder strength is denoted by $\epsilon _{i}$ in the
%%Hamiltonian (\ref{H}), which is randomly distributed within $[-W/2,W/2]$.
Fig. \ref{fig:level} shows the results for a chosen disorder configuration
with $W=0.05$ in a $32\times 32$ lattice. A careful examination of the
energy levels reveals that each eigenstate goes up and down, making several
large angle turns due to backward scattering. These energy levels never
cross with each other, except for at $\Phi=0$ and $\pi$, where two levels
become exactly degenerate (Kramers degeneracy). Then if one follows any
Kramers degenerated pair of states starting from $\Phi=0$ to $\Phi=2\pi$,
one will always go back to exactly the same pair of states. Namely, after
the adiabatic insertion of one flux quantum, all states evolve exactly back
to the starting states.

\begin{figure}[t]
\resizebox{80mm}{!}{    \includegraphics{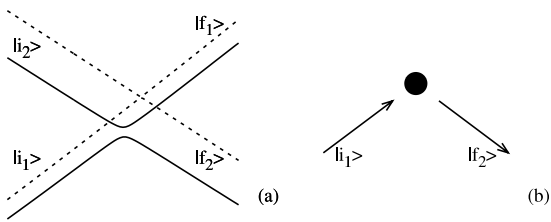} } \centering
\caption{(a) The comparison of the evolutions of two levels for clean system
(dashed line) and disordered system with $W = 0.05$ (dashed line). The
levels are crossing at clean system and anti-crossing at disorder system.
The $|i_1 \rangle$ and $|i_2 \rangle$ are the initial states while the $|f_1
\rangle$ and $|f_2 \rangle$ are the final states, see the text for details.
(b) The process of a state $|i_1 \rangle$ scattered to $|f_2 \rangle$ by the
disorder. Such a process is leading to the anti-crossing as shown in (a). }
\label{fig:scattering}
\end{figure}

For comparison with the $W=0$ case, in Fig. \ref{fig:scattering}(a) we
combine an enlarged plot of two adjacent energy levels in Fig.\ref%
{fig:scattering} together with the pure case, where the solid lines are for
the disordered case and the dashed lines are for the clean case. In the pure
case, an electron at the state $|i_{1}\rangle $ ($|i_{2}\rangle $) will
finally evolve to $|f_{1}\rangle $ ($|f_{2}\rangle $) with increasing flux
due to the level crossing. However, in contrast to the level crossing in the
pure system, two energy levels, $|i_{1}\rangle ,|i_{2}\rangle ,$ in the
disordered system generally show a level repulsion as one increases the flux 
$\Phi ,$ to evolve into the final states of $|f_{2}\rangle $ and $%
|f_{1}\rangle ,$ respectively [see Fig. \ref{fig:scattering}(a)]. The level
repulsion in the disordered case represents a scattering process shown in
Fig. \ref{fig:scattering}(b).

% which can be labelled by three quantum numbers:
% $n,\lambda ,P_{x}$ with the wavefunction given by $e^{iP_{x}}u_{n\lambda
%   k_{x}}(y)$ and the energy $E_{n\lambda }(P_{x})$ forming 1D bands with
% $n=1,2,\cdots ,2L$ (for the open BC along the $y$-axis) and the
% \textquotedblleft helicity\textquotedblright\ $\lambda =\pm 1$ (in replacement
% of the spin index) as shown by the solid curves in the inset of Fig.
% \ref{fig:energy_spectrum}.  $P_{x} = 2 k \pi/L + \theta / L$ is the momentum
% along $x$ direction which is the lattice momentum at $\theta = 0$ and
% increases for $k > 0$ and decreases for $k < 0$ with increasing the flux,
% shown as the arrows in fig. \ref{fig:energy_spectrum}.  The helicity $\lambda$
% is unchanged under time reversal(T) transformation and tracks the sign of the
% contribution to the spin current.

% \begin{figure}[t]
%   \resizebox{80mm}{!}{    \includegraphics{band_o.eps} } \centering
%   \caption{The inset is the Energy spectrum of a clean sample at $\protect%
%     \theta =0$. The main panel shows two of these bands. The two arrows
%     illustrate the evolution of the two states with tuning the parameter $%
%     \protect\theta $, which start from the beginning of the arrows at
%     $\protect%
%     \theta =0$ and arrive at the end of the arrows at $\protect\theta
%     =2\protect%
%     \pi $(see the text for details). The dashed line indicate that the Fermi
%     energy.}
%   \label{fig:energy_spectrum}
% \end{figure}

We have done similar calculations at different lattice sizes, such as $%
8\times 8$, $16\times 16,$ and $24\times 24$. The results are all similar
except that the ``gaps'' characterizing the level repulsion decrease with
the lattice size, which are roughly proportional to $1/N$. According to the
discussion in Sec. \ref{sec:gelc}, the adiabatic insertion of the flux $\Phi 
$ will not generate true spin transfer or accumulation as the ground state
simply evolves back after a flux quantum insertion. In fact, we have checked
various weak disorder strengths and always found the same generic result.

Therefore, we conclude that there is no intrinsic spin Hall accumulation
(spin transfer) for the 2DHG system in the disorder case. 
%%%similar to the case of the 2D Rashba model.\cite{kubo2}
This conclusion is in sharp contrast to the calculated SHC at small $W$ in
Sec. \ref{sec:shc}, which remains finite and close to the value of the pure
case when the disorder is weak. According to the discussion at the beginning
of Sec. \ref{sec:sa}, these two approaches, for a bulk system on a torus or
as an open system with two edges, are not necessarily in accordance with
each other, because the spin current continuity equation is broken. For
example, a backward scattering of an electron by an impurity shown by Fig. %
\ref{fig:scattering}(b) can simultaneously change the sign of its momentum 
\emph{as well as} its spin direction due to the SOC. Then one can find that
the spin current is still additive (same sign) in this scattering process,
which contributes to the SHC accordingly, but there should be no true spin
transfer taking place because of the backward scattering of the spin carrier
-- electron which moves back in direction after the scattering.

\subsection{Edge state}

\label{sec:dis}

\begin{figure}[t]
\resizebox{70mm}{!}{    \includegraphics{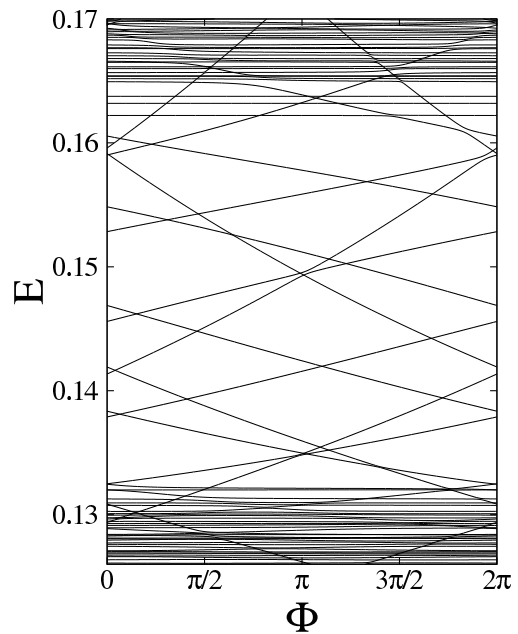} } \centering
\caption{The evolution of the single electron states with $\Phi$ in the
system with an external magnetic field on a $32 \times 32$ lattice. The
strength of the magnetic field is $\protect\pi/8$ per plaquette.}
\label{fig:mag_level}
\end{figure}

To further understand the condition for the presence of level crossing in
the disordered case, we apply an external magnetic field perpendicular to
the 2D sample (with a flux strength $\pi /8$ per plaquette) and repeat the
above calculation. We find that most of the ``gaps'' still behave as $1/N$ ,
similar to those in the non-magnetic-field case. These are bulk states. But
we can also identify a few levels whose level-repulsion ``gaps'' reduce
exponentially with $N$ with the increase of the sample size. For example,
such kind of ``gap'' at $32\times 32$ lattice size is about $1/30$ of the
one at $16\times 16$ lattice size, an exponential dependence on system
length. 
%instead of reducing only about $1/4$ for a level-repulsion ``gap''.
We show some of these ``level crossings'' in Fig. \ref{fig:mag_level} with
the same parameters as in Fig. \ref{fig:level}. The ``gaps'' in Fig. \ref%
{fig:mag_level} are much smaller than the ones in Fig. \ref{fig:level} and
hard to distinguish with the naked eye. %A careful examination clearly
The exponential decaying form of the gaps indicate the existence of level
crossings in the thermodynamic limit in the presence of a perpendicular
magnetic field, which corresponds to an IQHE. Then a net charge (spin) is
transported from one edge to the other correspondingly.\cite{kubo2}

We have also systematically compared the spatial distribution of electron
density $\rho _{i}=\langle c_{i}^{\dagger }c_{i}\rangle $ of the states in
the 2DHG system with and without the presence of a perpendicular magnetic
field. %%%Fig. \ref{fig:dis} shows three
%%representative examples, with the top one for a state in the non magnetic
%%case, the middle one for a state which anticrosses with others in the
%%magnetic case, and the bottom one as a state which crosses with some other
%%states in the magnetic case. The top one shows an extended state for the SOC
%system. The middle one is a localized state due to the Anderson
%%localization. And the bottom one is an edge state which is localized at the
%%edge along the $\hat{y}$-direction but remains extended along the $\hat{x}$%
%%-direction.
It is found that the level crossing only occurs for the edge states.
Physically this is because in the thermodynamic limit the distance between
two edge states in the opposite sides of the sample goes to infinity, which
makes the scattering between them vanishes exponentially. Thus the absence
of level crossing (and the intrinsic spin Hall transfer) in the 2DHG system
can be attributed to the fact that there do not exist any edge states in
such a system in the absence of perpendicular external magnetic fields.

\section{Conclusions and Discussions}

\label{sec:dac}

In the present work, we have numerically calculated the SHC through the Kubo
formula at different lattice sizes and impurity strengths. A reasonable
finite-size scaling analysis shows that the bulk SHC remains finite in the
thermodynamic limit in the disordered 2DHG up to a critical disorder
strength $W_{c}=5$. This result is consistent with the perturbative
calculation based on the vertex correction.\cite{2dhg}

However, we have also found the strong evidence that the net spin Hall
accumulation or spin transfer disappears in the disordered case based on a
numerically performed Laughlin's gauge experiment. 
% to probe the spin accumulation at the open boundaries of a
%belt-shaped sample upon an adiabatical insertion of a flux quantum.
We have interpreted the conflicting results of the SHC and direct spin
accumulation calculations as due to the fact that the spin is not conserved
in an SOC system and there is no longer a conservation law to govern the
spin current. Therefore the SHC is no more a unique quantity to characterize
the spin transport, where electron spin can relax and vanish in the bulk
without being transported across the sample by a transverse electric field.

By comparing with the case in the presence of an applied perpendicular
magnetic field, which is in the IQHE regime, we have shown that the absence
of an intrinsic spin transfer (accumulation) can be related to that there
are no edge states in the 2DHG system, in contrast to the IQHE case.

We further address the validity of the above adiabatic argument in the
thermodynamic limit. It is noted that since an intrinsic spin transfer
(accumulation) is always \emph{absent} in a finite size system according to
the above argument, it is very hard to imagine that such an effect could be
restored as an \emph{intrinsic} one by the Landau-Zener tunnelling \cite%
{kubo1} between two anticrossing levels in the thermodynamic limit. Of
course, the Landau-Zener tunnelling between the level repulsion gaps in the
large sample size limit (the level repulsion gap vanishes in a $1/N$
fashion, with the \emph{same }rate as the average level spacing vanishing in
this limit) can still contribute to \emph{dissipative} transport currents,
and similar to the conventional charge transport, one may expect a
dissipative spin transport term appearing, besides the null \emph{intrinsic}
spin transfer. Here the dissipative spin transport is qualitatively
different from the true level crossing with edge states. At a finite $W$, in
the longitudinal channel, the so-called \textquotedblleft Thouless
conductance\textquotedblright\ term can be obtained based on the sensitivity
of the energy level to the threading flux, giving rise to a finite $\sigma
_{xx}$,\cite{ando} which is subject to the detailed scattering mechanism as
contributed by the levels near the Fermi energy. In the present SOC system,
a dissipative spin Hall transfer related to this longitudinal conductance is
indeed possible, which similarly involves only the states near the Fermi
energy. We note that such a dissipative term may also be important for the
spin-polarization measured experimentally\cite{exp1,exp2}.

The vanishing spin transfer revealed through Laughlin's Gedanken gauge
experiment for the present 2DHG and 2D electron Rashba model\cite{kubo2}
under an adiabatic procedure may be a generic behavior for a metallic system.%
\cite{haldane} It would be interesting to directly probe this property for
more SOC systems, including the 3D Luttinger model, although numerically it
is more challenging for the latter as it is much harder to do the
finite-size scaling for the level repulsion gaps in a 3D system. Finally, it
is interesting to note that several recent papers have studied topological
SHE in insulating electron systems with SOC.\cite{tshe} In these models, it
becomes clear that bulk intrinsic SHC can result in spin transfer and
accumulation, with the presence of spin-polarized edge states, which is
associated with a topological invariant. This is in agreement with our
general conclusion that the intrinsic spin transfer relies on the existence
of current-carrying edge state.

\begin{acknowledgments}
\textbf{Acknowledgment:} We would like to thank F.D.M. Haldane, S.C. Zhang,
and M.W. Wu for insightful discussions. This work is supported by NSFC
grants 10374058 and 90403016 (ZYW), ACS-PRF 41752-AC10, the NSF
grant/DMR-0307170 (DNS). The computation of this project was partially
performed on the HP-SC45 Sigma-X parallel computer of ITP and ICTS, CAS.
\end{acknowledgments}

\end{document}